\documentstyle[preprint,aps,fleqn]{revtex} % PREPRINT format

\newfont{\bi}{cmbxti10 scaled\magstep2}
\begin{document}
\draft
%**end of header

\title{{\bi Ab initio} study of the volume dependence of
dynamical and thermodynamical properties of silicon.}
\author{G.-M. Rignanese, J.-P. Michenaud, and X. Gonze.}
\address{Unit\'e de Physico-Chimie et de Physique des
Mat\'eriaux,
Universit\'e Catholique de Louvain,\\
1 Place Croix du Sud, B-1348 Louvain-la-Neuve, Belgium}
\date{\today}
\maketitle

\begin{abstract}
Motivated by the negative thermal expansion observed for silicon
between 20~K and 120~K,
we present first an {\it ab initio} study of the volume dependence
of interatomic force constants,
phonon frequencies of TA($X$) and TA($L$) modes,
and of the associated mode Gr\"uneisen parameters.
The influence of successive nearest neighbors shells is analysed.
Analytical formulas,
taking into account interactions up to second nearest neighbors,
are developped for phonon frequencies of TA($X$)
and TA($L$) modes and the corresponding mode Gr\"uneisen parameters.
We also analyze the volume and pressure dependence
of various thermodynamic properties (specific heat, bulk modulus,
thermal expansion), and point out the effect of the negative mode
Gr\"uneisen parameters of the acoustic branches on these properties.
Finally, we present the evolution of the mean square atomic displacement and
of the atomic temperature factor with the temperature for different volumes,
for which the anomalous effects are even greater.
\end{abstract}

%\vskip 5mm

% \begin{multicols}{2}[]

%Introduction
\section*{Introduction}
In the past few years, theoretical and algorithmic advances
have made it possible to determine the thermodynamical properties of solids
(such as specific heat or thermal-expansion coefficient)
from first principles calculations
\cite{Note_thermodynamic,Lee-Gonze95}.

Silicon is a choice system for testing these methods,
since accurate measurements on high-purity samples exist
over a wide range of temperature.
Moreover, it presents a negative thermal-expansion coefficient
between 20~K and 120~K which is of fundamental interest.
This unusual thermal-expansion behavior can be
attributed to the negative Gr\"uneisen parameters
(i.e., phonon frequency increases as crystal volume increases)
of the transverse acoustic (TA) phonons near the Brillouin-zone boundary.
It is interesting to note that this anomalous behaviour also affects
other properties of silicon, such as the mean square atomic displacement
(as shown later in this paper).

This anomalous negative thermal expansion has been analyzed in previous
studies,
most of them relying on the quasiharmonic approximation \cite{Boyer79}.
In this approach, the volume dependence of phonon frequencies
must be determined.
Kagaya, Shuoji, and Soma \cite{Kagaya88} used a perturbation
treatment with a model pseudopotential to calculate the specific heat and the
thermal-expansion coefficient.
Biernacki and Scheffler \cite{Biernacki-Scheffler89}
employed a Keating model with two parameters,
which were extracted from local-density pseudopotential calculations,
to compute the thermal-expansion coefficient.
Fleszar and Gonze \cite{Fleszar-Gonze90} performed a model-free computation of
the thermal expansion, using a linear-response technique
within DFT \cite{Baroni87}.
Xu {\it et al.} \cite{Xu91}
used a tight-binding model to calculate the thermal-expansion coefficient and
mode Gr\"uneisen parameters;
they also showed by using a simple model that the negative values of the
mode Gr\"uneisen parameter could be attributed to a larger contribution
from the central part of forces than the angular part of forces.
More recently, Wei, Li, and Chou \cite{Wei94} extracted
interatomic force constants from {\it ab initio} calculations
of planar forces \cite{Wei-Chou94},
and calculated the specific heat, the overall Gr\"uneisen parameter,
and the thermal-expansion coefficient.

In this paper, we use a variational approach
to density-functional perturbation theory
to calculate volume-dependent dynamical properties of silicon
(see Sec.~\ref{sec:dyn-prop}).
We present first an {\it ab initio} study of the volume dependence
of interatomic force constants up to twenty-fifth nearest neighbors.
Wei and Chou had also presented such a calculation but only for one volume and
up to eighth nearest neighbors \cite{Wei-Chou94,Wei-Chou92}.
Phonon frequencies of TA($X$) and TA($L$) modes,
and of the associated mode Gr\"uneisen parameters are also calculated
for different volumes.
The influence of successive nearest neighbors shells is analysed.
We confirm that the contribution of atoms connected by zig-zag chain along
[1~1~0] direction are the most important, as suggested by Mazur and Pollmann
\cite{Mazur-Pollmann89}.
But we show that the contributions of fifth, sixth, and seventh atoms along the
chain (respectively thirteenth, eightenth and twenty-fifth nearest neighbors)
are not negligible contrarily to what had been speculated by Wei and Chou
\cite{Wei-Chou94}.
Analytical formulas,
taking into account interactions up to second nearest neighbors,
are developped for phonon frequencies of TA($X$)
and TA($L$) modes and the corresponding mode Gr\"uneisen parameters.
In Sec.~\ref{sec:thm-prop}, we analyze the volume and pressure dependence
of various thermodynamic properties (specific heat, bulk modulus,
thermal expansion).
The effect of zero-point motion on static equilibrium properties
(lattice constant and bulk modulus) also analyzed.
We point out the effect of the negative mode
Gr\"uneisen parameters of the acoustic branches on these properties.
In Sec.~\ref{sec:atom-fact}, we present the evolution
of the mean square atomic displacement and
of the atomic temperature factor with the temperature for different volumes.
Anomalous effects present at all temperature are investigated,
using a band-by-band decomposition.
Finally, we present our conclusions in the last section.

\section{Dynamical properties}
\label{sec:dyn-prop}
\subsection{Interatomic force constants}
\label{subsec:ifc}
To obtain accurate interatomic force constants (IFC's),
we first calculate from first principles the dynamical matrices
for 10 different wavevectors
in the irreducible Brillouin zone (IBZ)
(the special points mentioned in Ref.~\cite{Chadi-Cohen73})
using a variational approach to density-functional perturbation theory
Then, we perform a Fourier Transform of these
dynamical matrices in order to get the IFC's,
following Ref.~\cite{Note_thermodynamic,Gonze94}.
This sampling of dynamical matrices allows to obtain IFC's
in a real space box containing 512 atoms.
This is more accurate than what had been done in Ref.~\cite{Fleszar-Gonze90}
where the IFC's had only been calculated for 2 different wavectors
in the IBZ which corresponds to a real space box of 64 atoms.
It is also better than the work of Wei and Chou \cite{Wei-Chou94,Wei-Chou92}
who had included interatomic interactions up to eighth nearest neighbors,
which corresponds to 99 atoms.

Our calculations, performed
within the local density approximation (LDA) \cite{Kohn-Sham65}, use a
preconditioned conjugate gradient algorithm \cite{Teter89,Payne92}.
We use a rational polynomial parametrization of the
exchange-correlation energy functional \cite{Teter}, which is based
on the Ceperley-Alder gas data \cite{Ceperley-Alder80}.
The electronic wavefunctions are sampled on a mesh of 10 special $k$ points
in the IBZ and expanded in terms of a set of plane-waves
whose kinetic energy is limited to 10 Hartree.
The ``all-electron'' potentials are replaced by an {\it ab initio}, separable,
norm-conserving pseudopotential built following the scheme proposed
in Ref.~\cite{Hamann89}.
The calculated equilibrium lattice constant is 10.18 Bohr~\cite{Rignanese95}
(the influence of the zero-point motion on this value will be discussed
in Sec.~\ref{sec:thm-prop}),
whereas the experimental one is 10.26 Bohr.

The calculation of IFC's is performed for three different volumes,
corresponding to lattice constants $a$ of 10.00, 10.18, and 10.26 Bohr
respectively.
The IFC's for these different lattice constants
are listed in Table~\ref{tab:ifc}, in which coordinates
are in units of $a/4$ and notations for the force constants follow
Ref.~\cite{Herman59}.

As already noticed in Refs.~\cite{Wei-Chou94,Mazur-Pollmann89},
IFC's for atoms connected
by zig-zag chain along [1~1~0] direction (or the directions related to it
by symmetry) are the most important.
The magnitude of IFC's along these peculiar directions decays quite slowly.
The contribution of the first nearest neighbors ($\alpha_1$ and $\beta_1$)
is about 4 to 5 times smaller than on-site contribution ($\alpha_0$).
Going up to second nearest neighbours reduces the IFC's by another factor of 5
($\lambda_2$) or 10 ($\mu_2$ and $\nu_2$).
The contribution $\lambda_{25}$ of the seventh atoms along the chain
(coordinates (7,7,1) in units of $a/4$),
which are already the twenty-fifth nearest neighbors,
is still of the order of 1 percent of the contribution of the first nearest
neighbors, and of 5 percent of that of the second ones.
Comparatively, the biggest contribution of the seventh nearest neighbors
(these do not belong to the chain, but are twice closer than
the seventh atoms along the chain) is about 2 times smaller.

In order to vizualize these IFC's, we compute
the forces induced along the zig-zag chain
by the displacement of a generic atom of this chain in three
perpendicular directions (along
[1~1~0] --- parallel to the chain direction,
along [0~0~1] --- perpendicular to the chain direction, but in the
plane of the chain, and
along [1~$\bar 1$~0] --- perpendicular to the chain direction,
but out of the plane of the chain),
all the other atoms being kept fixed.
These forces can easily be obtained from IFC's by:
\begin{equation}
F_{\kappa\alpha}=-\Phi_{\alpha\beta}(\kappa;\kappa')
\left( R-R_0 \right)_{\kappa'\beta}
\end{equation}
where ${\bf R}-{\bf R}_0$ is the displacement of the generic atom, and
the IFC $\Phi_{\alpha \beta} (\kappa;\kappa')$ is the force exerted
on ion $\kappa$ in the direction $\alpha$
due to the displacement of ion $\kappa'$ in the direction $\beta$.

The resulting forces have been reproduced in Fig.~\ref{fig:chain}.
The directions of the forces induced by the displacement of the generic atom
in the [1~1~0], and [0~0~1] directions
show that the bonds along the chain will tend to bend
in order to keep constant the bond angles : at sufficient distance,
the forces are nearly perpendicular to the last connecting bond.
The decay of the magnitude of these forces along the chain is similar to that
of the IFC's.
At the contrary, the forces induced by the displacement in the [1~$\bar 1$~0]
direction are much more smaller.
Indeed, they have the generic form $(\mu-\nu,\nu-\mu,0)$
where $\mu$ and $\nu$ are of the same order of magnitude
(see Table~\ref{tab:ifc}).
It is thus more difficult to interpret them.

\subsection{TA($X$) and TA($L$) phonon frequencies and
associated mode Gr\"uneisen parameters}
\label{subsec:phon}
In previous studies, the negative mode Gr\"uneisen parameters associated with
the modes of the acoustic branches, near the zone boundaries,
has been shown to drive the negative thermal expansion.
We now focus on the analysis of the high symmetry
TA($X$) and TA($L$) modes because analytical results can be obtained for them.

The phonon frequencies
can be calculated by solving the dynamical equation:
\begin{equation}
\sum_{\kappa' \beta} \Phi_{\alpha \beta} (\kappa;\kappa')
u_\sigma(\kappa'\beta)=
M_\kappa \omega^{2}_{\sigma}
u_\sigma(\kappa \alpha)
\label{eq:dynamical}
\end{equation}
where $u_\sigma(\kappa \alpha)$ is the displacement in the
direction $\alpha$ of ion $\kappa$ for the normal mode $\sigma$,
$M_\kappa$ is the mass of ion $\kappa$,
and $\omega_\sigma$ is the frequency of the normal mode $\sigma$.

In order to analyse the influence of the successive nearest neighbors shells
for TA($X$) and TA($L$) modes,
we artificially limit the interactions taken
into account in the IFC matrix $\Phi_{\alpha \beta} (\kappa;\kappa')$
to the atoms whose distance is less than a cut-off radius.
Then, we increase the cut-off radius shell by shell until the sphere contains
all the atoms included in the real space box defined
by our sampling of the Brillouin zone.

In truncating the interaction to different shells,
we break the acoustic sum rule
(which states that moving all atoms as whole should not create any force)
unless it is reimposed by artificially changing the on-site IFC $\alpha_0$.
We analyzed both the results obtained with and without reimposing the acoustic
sum rule and found no significative difference regarding the convergence with
respect to the number of nearest neighbors shells included.
Thus, in the following, we will only reproduce the results obtained without
reimposing the acoustic sum rule.

The results are reproduced in Tables~\ref{tab:phonX} and \ref{tab:phonL}.
The well-converged calculated frequencies of TA($X$) and TA($L$) modes
are respectively
140.46~cm$^{-1}$ and 108.626~cm$^{-1}$ at the calculated equilibrium lattice
constant, 147.37~cm$^{-1}$ and 112.86~cm$^{-1}$ at the experimental one.
These values are in very good agreement with experiment \cite{Dolling}:
149.77~cm$^{-1}$ and 114.41~cm$^{-1}$.

The associated mode Gr\"uneisen parameters are calculated by:
\begin{equation}
\gamma_\sigma=-\frac{d\ln \omega_\sigma} {d\ln V}
\label{eq:gruneisen}
\end{equation}
where $\sigma$ is the index of the mode, and $V$ is the molar volume
($N_A$ $\times$ volume of the unit cell),
linked to the lattice constant by:
\begin{equation}
V=N_A \frac{a^3}{4}.
\end{equation}
The results, reproduced in Tables~\ref{tab:phonX} and \ref{tab:phonL},
are very sensitive to the volume variations.
The calculated mode  Gr\"uneisen parameters of TA($X$) and TA($L$) modes
are respectively -2.295 and -1.814 at the calculated equilibrium lattice
constant.
These values are not in very good agreement with experiment \cite{Weinstein75}:
-1.4 and -1.3.
However, if we consider the values -1.782 and -1.451 obtained
at the experimental equilibrium lattice constant,
the agreement is much better.

Considering for reference the values obtained by taking into account
all the atoms included in the real space box,
we see that the contribution of atoms connected by zig-zag chain
along [1~1~0] direction are the most important for the convergence of
the phonon frequencies of TA($X$) and TA($L$) modes,
and of the associated mode Gr\"uneisen parameters.
Beyond the fourth atom along the [1~1~0] chain
(eighth nearest neighbors),
the contribution of successive nearest neighbors not belonging to the chains
is not significant.
But the contributions of fifth, sixth, and seventh atoms along the chain
(respectively thirteenth, eightenth and twenty-fifth nearest neighbors)
are not negligible, contrarily to what had been speculated
by Wei and Chou \cite{Wei-Chou94}.
Including still further atoms along the chains would necessitate to use another
sampling of the dynamical matrix wavevectors in order
to get a bigger real space box.
Tables~\ref{tab:phonX} and \ref{tab:phonL} also show that, once interactions
with the first nearest neighbors are included, the anomalous behaviour of TA
modes at $X$ and $L$ points (negative Gr\"uneisen parameters) are reproduced.
This reinforce the pertinence of the model proposed in Ref.~\cite{Xu91}.

For TA($X$) and TA($L$) modes,
the eigenmodes $u_\sigma(\kappa \alpha)$ are known for any ion
from symmetry considerations (see Fig.~\ref{fig:displace}).
It is thus even possible to obtain
an analytical expression of phonon frequencies, from the knowledge
of the IFC's.

If we only consider on-site interaction
(just atom $\kappa$) in the sum of the left-hand side of
Eq.~(\ref{eq:dynamical}), we get:
\begin{equation}
\omega=\sqrt{\frac{\alpha_0}{m}}
\end{equation}
for both TA($X$) and TA(L) modes, where $m$ is the mass of silicon ion.
The associated mode Gr\"uneisen parameters can easily be obtained by inserting
this result in Eq.~(\ref{eq:gruneisen}):
\begin{equation}
\gamma=\gamma(\alpha_0)
\end{equation}
where $\gamma(\alpha_0)$ is the ``force Gr\"uneisen parameter'',
that we define as:
\begin{equation}
\gamma(\alpha_0)=-\frac{1}{2}\frac{d\ln \alpha_0} {d\ln V}.
\label{eq:force-grun}
\end{equation}
The force Gr\"uneisen parameters have been listed in Table~\ref{tab:ifc} for
for all IFC's whose magnitude is higher than $10^{-4}$ Hartree/Bohr$^3$.

If we consider interactions up to first nearest neighbors atoms, we get:
\begin{equation}
\omega_{TA(X)}=\sqrt{\frac{\alpha_0+4\beta_1}{m}}
\label{eq:phon1x}
\end{equation}
for TA($X$) mode, and
\begin{equation}
\omega_{TA(L)}=\sqrt{\frac{\alpha_0+2\alpha_1+2\beta_1}{m}}
\label{eq:phon1l}
\end{equation}
for TA(L) mode.
In fact, whatever the number of nearest neighbors taken into account,
the phonon frequencies will always be written as the square root of the ratio
of a linear combination of the IFC's and the mass.
The associated mode Gr\"uneisen parameters can be written as {\it weighted sum}
of the force Gr\"uneisen parameters corresponding to the IFC's under the root
in
Eqs.~(\ref{eq:phon1x}) and (\ref{eq:phon1l})~. For example,
for the TA($X$) mode,
inserting Eq.~(\ref{eq:phon1x}) in Eq.~(\ref{eq:gruneisen}), we get:
\begin{equation}
\gamma_{TA(X)}=
-\frac{V}{\sqrt{\alpha_0+4\beta_1}} \frac{d\sqrt{\alpha_0+4\beta_1}}{dV}=
-\frac{1}{2} \frac{V}{\alpha_0+4\beta_1} \frac{d\alpha_0+4\beta_1}{dV}.
\end{equation}
This can be rewritten as follows:
\begin{equation}
\gamma_{TA(X)}=
-\frac{1}{2} \left( \frac{\alpha_0}{\alpha_0+4\beta_1} \right)
\frac{V}{\alpha_0} \frac{d\alpha_0}{dV}
-\frac{1}{2} \left( \frac{4\beta_1}{\alpha_0+4\beta_1} \right)
\frac{V}{\beta_1} \frac{d\beta_1}{dV}.
\end{equation}
The definition of force
Gr\"uneisen parameters, Eq.~(\ref{eq:force-grun}), leads to
\begin{equation}
\gamma_{TA(X)}=
\left( \frac{\alpha_0}{\alpha_0+4\beta_1} \right) \gamma(\alpha_0)+
\left( \frac{4\beta_1}{\alpha_0+4\beta_1} \right) \gamma(\beta_1)
\label{eq:gruneisen1x}
\end{equation}
for TA($X$) mode, and
\begin{eqnarray}
\gamma_{TA(L)} &=&
\left( \frac{\alpha_0}{\alpha_0+2\alpha_1+2\beta_1} \right) \gamma(\alpha_0)+
\left( \frac{2\alpha_1}{\alpha_0+2\alpha_1+2\beta_1} \right) \gamma(\alpha_1)+
\nonumber \\
& &
\left( \frac{2\beta_1}{\alpha_0+2\alpha_1+2\beta_1} \right) \gamma(\beta_1)
\label{eq:gruneisen1l}
\end{eqnarray}
for TA(L) mode.

Interestingly the same argument holds, whatever the number of
nearest neighbors taken into account.
And thus, the same kind of formulas will be obtained for phonon frequencies and
associated mode Gr\"uneisen parameters.

Indeed, if we consider interactions up to second nearest neighbors atoms,
we get:
\begin{equation}
\omega_{TA(X)}=\sqrt{\frac{\alpha_0+4\beta_1-4\lambda_2}{m}}
\label{eq:phon2x}
\end{equation}
for TA($X$) mode, and
\begin{equation}
\omega_{TA(L)}=\sqrt{\frac{\alpha_0+2\alpha_1+2\beta_1-4\nu_2}{m}}
\label{eq:phon2l}
\end{equation}
for TA(L) mode.
The associated mode Gr\"uneisen parameters can be written as weighted sum of
the force Gr\"uneisen parameters corresponding to the IFC's under the root in
Eqs.~(\ref{eq:phon2x}) and (\ref{eq:phon2l}):
\begin{eqnarray}
\gamma_{TA(X)} &=&
\left( \frac{\alpha_0}{\alpha_0+4\beta_1-4\lambda_2} \right) \gamma(\alpha_0)+
\left( \frac{4\beta_1}{\alpha_0+4\beta_1-4\lambda_2} \right) \gamma(\beta_1)+
\nonumber \\
& &
\left( \frac{-4\lambda_2}{\alpha_0+4\beta_1-4\lambda_2} \right)
\gamma(\lambda_2)
\label{eq:gruneisen2x}
\end{eqnarray}
for TA($X$) mode, and
\begin{eqnarray}
\gamma_{TA(L)} &=&
\left( \frac{\alpha_0}{\alpha_0+2\alpha_1+2\beta_1-4\nu_2} \right)
\gamma(\alpha_0)+
\left( \frac{2\alpha_1}{\alpha_0+2\alpha_1+2\beta_1-4\nu_2} \right)
\gamma(\alpha_1)+
\nonumber \\
& &
\left( \frac{2\beta_1}{\alpha_0+2\alpha_1+2\beta_1-4\nu_2} \right)
\gamma(\beta_1)+
\left( \frac{-4\nu_2}{\alpha_0+2\alpha_1+2\beta_1-4\nu_2} \right)
\gamma(\nu_2)
\label{eq:gruneisen2l}
\end{eqnarray}
for TA(L) mode.

It should be noted that the weights in Eqs.~(\ref{eq:gruneisen1x}) to
(\ref{eq:phon2l}) are not necessarily included in the interval [0,1]
and can thus be negative.
This explains why, though almost all force Gr\"uneisen parameters are positive
(see Table~\ref{tab:ifc}), a negative mode Gr\"uneisen parameter can be
obtained.

For example, in Eq.~(\ref{eq:gruneisen1x}), we get for $a$=10.18 Bohr:
\begin{equation}
\gamma_{TA(X)}=
3.08 \times 1.1034+\left( -2.08 \right) \times 1.9817=
-0.72~,
\end{equation}
which is just what is obtained in Table~\ref{tab:phonX}.
If all force Gr\"uneisen parameters were equal to 1, we would also get 1 for
the mode Gr\"uneisen parameter.
This shows that the origin of the negative Gr\"uneisen parameter is
the rather important difference between $\gamma(\beta_1)$ and
$\gamma(\alpha_0)$.

\section{Thermodynamic properties}
\label{sec:thm-prop}
Using the calculated phonon frequencies,
the temperature-dependent phonon contribution $\Delta F$ to the
Helmholtz free energy $F(V,T)$ is calculated as described
in Ref.~\cite{Lee-Gonze95}.
This contribution is added to the energy of the static lattice $E(V)$,
calculated previously \cite{Rignanese95},
to get $F(V,T)$ for a set of three volumes and for various temperatures.
This function is then interpolated as a function of $V$ by a second order fit.
The equilibrium volumes (or lattice constants)
at various temperatures are determined by minimizing
$F(V,T)$ as a function of $V$,
as shown in Fig.~\ref{fig:minf}.
{}From $F(V,T)$ any other thermodynamical property can be accessed.

We first analyze briefly the specific effect of the zero-point motion.
We find that the zero-point contribution to Helmholtz free energy F
is $\Delta F_0$=12~J/mol, which is about 2.5\% of the cohesive
energy (460~kJ/mol \cite{Davis61}).
This zero-point contribution causes the lattice constant to be shifted
from 10.1894~Bohr to 10.1974~Bohr, which is a change smaller than 0.1~\%,
and the bulk modulus $B_T$ from 1.0387~Mbar to 1.0292~Mbar,
which is a change of 1~\%.
Note however that the change in the bulk modulus includes two effects:
first, it is linked to the second derivative of $F(V)$ that includes the
zero-point motion contribution $\Delta F_0$;
and second, it is calculated for a different volume due to the shift of the
lattice constant.
In Fig.~\ref{fig:bt},
we present the temperature dependence of $B_T$ for different pressures.

The entropy can be calculated following Ref.~\cite{Lee-Gonze95}.
We get $S$(298.15~K)=19.3~J~K$^{-1}$~mol$^{-1}$, to be compared
to the experimental value of 18.81~J~K$^{-1}$~mol$^{-1}$
\cite{Landolt-Bornstein}.
We have also calculated the variation
of enthalpy $H=F+T S-P V$ between 0 and 298.15~K.
We get 3.285~J~mol$^{-1}$ whereas the experimental value is 3.217~J~mol$^{-1}$
\cite{Landolt-Bornstein}.
We present in Fig.~\ref{fig:cpm}
the molar constant-pressure specific heat $C_{P,m}$ for various pressures.
This result agrees quite well with experimental work
\cite{Flubacher59,Desai86}.
It is interesting to note that, for temperature higher than 85~K,
$C_{P,m}$ is higher for low values of the pressure.
Whereas, for temperature lower than 85~K,
it is just the contrary.
This means that there exists a temperature around 85~K
for which $C_{P,m}$ is independent of pressure.
In order to understand this observation,
we use the following relation (see Ref.\cite{Callen}):
\begin{equation}
\left( \frac{\partial S}{\partial P} \right)_T =
- \left( \frac{\partial V}{\partial T} \right)_P
\end{equation}
to write:
\begin{equation}
\left( \frac{\partial C_{P,m}}{\partial P} \right)_T =
T \left( \frac{\partial^2 S}{\partial P \partial T} \right)=
-T \left( \frac{\partial^2 V}{\partial T^2} \right)_P=
-T V \left( \alpha^2_P +\left( \frac{\partial \alpha_P}{\partial T}\right)_P
\right).
\end{equation}
This shows that in order to have $C_{P,m}$ independent of pressure,
one must have a decreasing $\alpha_P$.
This is precisely the case of silicon between 20~K and 100~K, where the
thermal-expansion coefficient gets more and more negative.

This thermal-expansion coefficient is well reproduced
by our calculations (see Fig.~\ref{fig:alpha}).
It is worth noting that increasing the pressure reinforces
the anomalous negative behaviour of this property.
This is confirmed by the strong negative value of the overall Gr\"uneisen
parameter at high pressure (see Fig.~\ref{fig:gamma}).
This anomalous behaviour has also consequences on other properties.
We present in Fig.~\ref{fig:cpm-cvm}, the difference $C_{P,m}-C_{V,m}$ between
the molar constant-pressure specific heat $C_{P,m}$ and
the molar constant-volume specific heat $C_{V,m}$ for various pressures.
At high temperature $C_{P,m}-C_{V,m}$ increases with increasing pressure,
while at low temperature, it is the contrary and the curve presents a bump.
This is can be deduced from the following relation \cite{Callen}:
\begin{equation}
C_{P,m}-C_{V,m}=\frac{\alpha^2_P V T}{\kappa_T},
\end{equation}
where $\kappa_T=1/B_T$.
Similarly, the difference $B_S-B_T$ between
$B_S$ the bulk modulus calculated from $F$ at constant entropy and
$B_T$ the bulk modulus calculated from $F$ at constant temperature
presents the same behaviour as $C_{P,m}-C_{V,m}$
as shown in Fig.\ref{fig:bs-bt}.
Indeed, we can write \cite{Callen}:
\begin{equation}
\kappa_S-\kappa_T=\frac{\alpha^2_P V T}{C_{P,m}}
\end{equation}
where $\kappa_S=1/B_S$.

\section{Atomic temperature factor}
\label{sec:atom-fact}
At finite temperature $T$, the intensity of X-ray diffraction from
the crystal is reduced, due to atomic motion.
The atomic temperature factor $e^{-W(\kappa)}$ characterizes the oscillations
of
atom $\kappa$ around its equilibrium position.
It is defined \cite{Willis-Pryor} by:
\begin{equation}
e^{-W(\kappa)}=
\exp \left( -\frac{1}{2} \sum_{\alpha \beta} B_{\alpha \beta}(\kappa)
G_{\alpha} G_{\beta} \right)
\end{equation}
where $G_{\alpha}$ is the component of scattering wavevector {\bf G} which is a
reciprocal lattice vector,
and $B_{\alpha \beta}(\kappa)$ is the mean square atomic
displacement \cite{Lee-Gonze95}
of atom $\kappa$ along the
directions $\alpha$ and $\beta$:
\begin{equation}
B_{\alpha \beta}(\kappa)=
\frac{1}{N_A M_\kappa}
\sum_{{\bf q},l} \frac{\hbar}{2 \omega({\bf q},l)}
\coth \frac{\hbar \omega({\bf q},l)}{2 k_B T}
e_i(\kappa | {\bf q},l)
e^*_j(\kappa | {\bf q},l)
\label{eq:thm-fact}
\end{equation}
where $M_\kappa$ is the mass of atom $\kappa$,
and $e_i(\kappa | {\bf q},l)$ is the $i$th component of the eigenvector
associated with the mode l at ${\bf q}$ in the lattice coordinates.

In the case of silicon, there is only one kind of atom.
Thus, for all atoms $e^{-W(\kappa)}$ are identical.
The reduction of the diffusion intensity is given by
$e^{-2 W(\kappa)}$, which is usually called the Debye-Waller factor.
The symmetry of the diamond structure imposes that:
\begin{equation}
B_{\alpha \beta}(\kappa)=B \delta_{\alpha \beta}.
\end{equation}
(This $B$ is not to be confused with the bulk modulus.)

In Table~\ref{tab:thm-fact}, the results obtained for the mean square
atomic displacement,
at three different volumes, corresponding to lattice constants $a$ of 10.00,
10.18, and 10.26 Bohr respectively, are compared with experimental results.
The agreement is on the order of a few percent.

Interestingly, even at room temperature,
the mean square atomic displacement decreases with increasing volume,
contrary to the intuition.
This can easily be understood from the definition of $B$
(see Eq.~\ref{eq:thm-fact}).
Indeed, it is written as a sum over all phonon bands of a term which
is proportional to the inverse square of the frequency
of the mode $\omega^{-2}$
(since $\coth x \rightarrow x^{-1}$ for small $x$).
Thus, it is determined mostly by acoustic branches.
This is confirmed by a band-by-band decomposition of $B$
(see Table~\ref{tab:thm-fact})
where the two lowest bands account for more than $2/3$ of $B$.
As the TA band and the first LA band
exhibit a negative mode Gr\"uneisen parameter,
we see that their contribution to $B$ decreases with increasing volume,
at the contrary of the contribution of the second LA band and
optic bands.

The thermal parameter $B$ and the atomic temperature factor $e^{-W(\kappa)}$,
for diffraction with scattering vector ${\bf G}=(2 \pi/a_0) {\bf \hat z}$,
are also calculated as a function of temperature (see Figs.~\ref{fig:thm-fact}
and \ref{fig:atom-temp-fact}).
The atomic temperature factor is not 1 even at 0~K due to the zero-point
motion.

%Conclusions
\section*{Conclusions}
\label{sec:concl}
In this paper, dynamical properties of silicon have been calculated
using a variational approach to density-functional perturbation theory.
We have presented first an {\it ab initio} study of the volume dependence
of interatomic force constants up to twenty-fifth nearest neighbors.
Phonon frequencies of TA($X$) and TA($L$) modes,
and of the associated mode Gr\"uneisen parameters have also been calculated
for different volumes.
The influence of successive nearest neighbors shells has been analysed.
This study has confirmed that
the contribution of atoms connected by zig-zag chain along
[1~1~0] direction is the most important.
It has also proven that
the contributions of fifth, sixth, and seventh atoms along the
chain (respectively thirteenth, eightenth and twenty-fifth nearest neighbors)
are not negligible.
Analytical formulas,
taking into account interactions up to second nearest neighbors,
have been developped for phonon frequencies of TA($X$)
and TA($L$) modes and the corresponding mode Gr\"uneisen parameters.
The volume and pressure dependence
of various thermodynamic properties (specific heat, bulk modulus,
thermal expansion) were also analyzed.
We have pointed out the effect of the negative mode
Gr\"uneisen parameters of the acoustic branches on these properties.
The effect of zero-point motion was also investigated.
Finally, we have presented the evolution
of the mean square atomic displacement and
of the atomic temperature factor with the temperature for different volumes,
emphasizing the anomalous effects due to the negative mode Gr\"uneisen
parameters, present at all investigated temperatures.

%Acknowledgments
\acknowledgments
We thank J.-M. Beuken for permanent computer assistance.
The ground state results were obtained using a version of the software program
Plane\_Wave (written by D. C. Allan),
which is marketed by Biosym Technologies of San Diego.
Two of the authors (G.-M. R., and X. G.) have benefited from
financial support of the National Fund for Scientific Research (Belgium).
This paper presents research results of Belgian Program on
Interuniversity Attraction Poles initiated by the Belgian State, Prime
Minister's Office, Science Policy Programming.
We also acknowledge the use of the RS 6000 workstation from the common
project between IBM Belgium, UCL and FUNDP.

%References

%Figures
\begin{figure}
\caption{Vibrational motion corresponding to the (a) TA($X$) mode and (b)
TA($L$) mode in silicon.
The displacements of ions are along the [1~1~0] direction in (a) and along
the [1~1~$\bar 2$] direction in (b).}
\label{fig:displace}
\end{figure}

\begin{figure}
\caption{Forces induced (linear response) along the [1~1~0] zig-zag chain
by the unit displacement of a generic atom (in grey) of this chain
in the (a) [1~1~0], (b) [0~0~1], and (c) [1~$\bar 1$~0] directions.
The unit vectors corresponding to these directions are respectively
{\bf \^x}, {\bf \^y},and {\bf \^z}.
The arrows starting from white atoms represent the direction of the forces,
while the absolute value of it is written close to it.
Forces along the [1~$\bar 1$~0] direction are indicated by dots, whereas
crosses refer to forces in the opposite direction.
The forces are expressed in Hartree/Bohr.}
\label{fig:chain}
\end{figure}

\begin{figure}
\caption{Volume dependence of the Helmholtz free energy $F$
for 4 different temperatures.
The smallest is 0~K (upper curve),
the biggest is 300~K (lower curve)
with an increment of 100~K between the curves.
The equilibrium volume $V_0(T)$ is located at the minimum of each curve.
Between 20~K and 120~K, this volume decreases due to the negative
thermal-expansion coefficient.}
\label{fig:minf}
\end{figure}

\begin{figure}
\caption{Temperature dependence of the theoretical bulk modulus $B_T$
for 4 different pressures.
The smallest is 0 Pa (lower curve),
the biggest is 6 10$^9$ Pa (upper curve)
with an increment of 2 10$^9$ Pa between the curves.
$B_T$ is expressed in Mbar.
Temperature is in~K.}
\label{fig:bt}
\end{figure}

\begin{figure}
\caption{Temperature dependence of
the constant-pressure specific heat $C_{P,m}$
for 2 different pressures.
The smallest is 0 Pa (upper curve at high temperature),
the biggest is 6 10$^9$ Pa (lower curve at high temperature).
$C_{P,m}$ is expressed in J mol$^{-1}$~K$^{-1}$.
The crosses indicate experimental data \protect \cite{Flubacher59}.
Temperature is in~K.}
\label{fig:cpm}
\end{figure}

\begin{figure}
\caption{Temperature dependence of the volumic thermal-expansion coefficient
$\alpha_{p}$ for 4 different pressures.
The smallest is 0 Pa  (upper curve),
the biggest is 6 10$^9$ Pa  (lower curve)
with an increment of 2 10$^9$ Pa between the curves.
$\alpha_{p}$ is expressed in~K$^{-1}$.
The crosses indicate experimental data \protect \cite{Landolt-Bornstein}.
Temperature is in~K.}
\label{fig:alpha}
\end{figure}

\begin{figure}
\caption{Temperature dependence of overall Gr\"uneisen parameter
$\gamma$ for 4 different pressures.
The smallest is 0 Pa  (upper curve),
the biggest is 6 10$^9$ Pa  (lower curve)
with an increment of 2 10$^9$ Pa between the curves.
The crosses indicate experimental data \protect \cite{Ibach69}.
Temperature is in~K.}
\label{fig:gamma}
\end{figure}

\begin{figure}
\caption{Temperature dependence of the difference between
the constant-pressure specific heat $C_{P,m}$ and
the constant-volume specific heat $C_{V,m}$
for 4 different pressures.
The smallest is 0 Pa  (lower curve at high temperature),
the biggest is 6 10$^9$ Pa  (upper curve at high temperature)
with an increment of 2 10$^9$ Pa between the curves.
$C_{P,m}$ is expressed in J mol$^{-1}$~K$^{-1}$.}
\label{fig:cpm-cvm}
\end{figure}

\begin{figure}
\caption{Temperature dependence of the difference between
$B_S$ the bulk modulus calculated from $F$ at constant entropy and
$B_T$ the bulk modulus calculated from $F$ at constant temperature,
for 4 different pressures.
The smallest is 0 Pa  (lower curve at high temperature),
the biggest is 6 10$^9$ Pa  (upper curve at high temperature)
with an increment of 2 10$^9$ Pa between the curves.
$B_S-B_T$ is expressed in Mbar.
Temperature is in~K.}
\label{fig:bs-bt}
\end{figure}

\begin{figure}
\caption{Temperature dependence of
mean square atomic displacement $B$ for silicon atoms
in bulk silicon at three different volumes, corresponding to
lattice constants $a$ of 10.00 (dashed line),
10.18 (solid line), and 10.26 Bohr (dotted line) respectively.
The values of $B$ are expressed in \AA$^{2}$.
Temperature is in~K.}
\label{fig:thm-fact}
\end{figure}

\begin{figure}
\caption{Temperature dependence of atomic temperature factor $e^{-W(\kappa)}$
for silicon atoms in bulk silicon for diffraction with the scattering vector
${\bf G}=(2 \pi/a_0) {\bf \hat z}$
at three different volumes, corresponding to
lattice constants $a$ of 10.00 (dashed line),
10.18 (solid line), and 10.26 Bohr (dotted line) respectively.
Temperature is in~K.}
\label{fig:atom-temp-fact}
\end{figure}

%Tables
\begin{table}
\caption{Interatomic force constant
matrix elements and associated force Gr\"uneisen parameters
for silicon at three different volumes, corresponding to
lattice constants $a$ of 10.00, 10.18, and 10.26 Bohr respectively.
The convention of Ref.~\protect \cite{Herman59} for labelling the matrix
elements has been followed.
One atom is at the origin, while the coordinates
of the second atom are expressed in units of $a/4$,
the stars indicate the atoms
belonging to the zig-zag chain along [1~1~0] direction.
The first column indicates the number of the shell $NN$ to which the
second atom belongs. The force Gr\"uneisen parameters of IFC matrix
elements that are smaller than 10$^{-4}$ Hartree/Bohr$^2$ are not mentioned.
The interatomic force constant matrix elements are expressed in
Hartree/Bohr$^2$.}
\label{tab:ifc}
\newpage
\renewcommand{\arraystretch}{.59}
\begin{tabular}{cclrrrlrrr}
$NN$ &Coordinate &              &$a$=10.00  &$a$=10.18  &$a$=10.26
        &                       &$a$=10.00  &$a$=10.18  &$a$=10.26 \\
\hline
0 &(0,0,0)$^*$   &$\alpha_0$    & 0.15629 & 0.13904 & 0.13201
         &$\gamma(\alpha_0)$    & 1.0766  & 1.1034  & 1.1047 \\
1 &(1,1,1)$^*$   &$\alpha_1$    &-0.03779 &-0.03385 &-0.03225
         &$\gamma(\alpha_1)$    & 1.0215  & 1.0336  & 1.0275 \\
  &              &$\beta_1$     &-0.02880 &-0.02348 &-0.02137
         &$\gamma(\beta_1)$     & 1.8264  & 1.9817  & 2.0338 \\
2 &(2,2,0)$^*$   &$\mu_2$       &-0.00203 &-0.00182 &-0.00173
         &$\gamma(\mu_2)$       & 0.9600  & 1.0682  & 1.1216 \\
  &              &$\nu_2$       &-0.00200 &-0.00178 &-0.00168
         &$\gamma(\nu_2)$       & 1.0879  & 1.1641  & 1.1951 \\
  &              &$\delta_2$    & 0.00120 & 0.00111 & 0.00108
         &$\gamma(\delta_2)$    & 0.6487  & 0.6997  & 0.7234 \\
  &              &$\lambda_2$   & 0.00445 & 0.00433 & 0.00428
         &$\gamma(\lambda_2)$   & 0.2476  & 0.2456  & 0.2437 \\
3 &(\=1,\=1,\=3) &$\mu_3$       & 0.00035 & 0.00032 & 0.00031
         &$\gamma(\mu_3)$       & 0.6911  & 0.8662  & 0.9617 \\
  &              &$\nu_3$       &-0.00033 &-0.00035 &-0.00036
         &$\gamma(\nu_3)$       &-0.6046  &-0.5696  &-0.5546 \\
  &              &$\delta_3$    & 0.00032 & 0.00029 & 0.00028
         &$\gamma(\delta_3)$    & 0.8368  & 0.9285  & 0.9736 \\
  &              &$\lambda_3$   & 0.00006 & 0.00006 & 0.00006
         &$\gamma(\lambda_3)$   & & & \\
4 &(0,0,4)       &$\mu_4$       &-0.00020 &-0.00020 &-0.00019
         &$\gamma(\mu_4)$       & 0.2788  & 0.2831  & 0.2843 \\
  &              &$\lambda_4$   &-0.00007 &-0.00005 &-0.00004
         &$\gamma(\lambda_4)$   & & & \\
5 &(3,3,1)$^*$   &$\mu_5$       &-0.00022 &-0.00020 &-0.00019
         &$\gamma(\mu_5)$       & 1.1211  & 1.1069  & 1.0819 \\
  &              &$\nu_5$       &-0.00032 &-0.00026 &-0.00024
         &$\gamma(\nu_5)$       & 1.8115  & 1.9104  & 1.9240 \\
  &              &$\delta_5$    & 0.00064 & 0.00058 & 0.00055
         &$\gamma(\delta_5)$    & 0.9110  & 0.9660  & 0.9882 \\
  &              &$\lambda_5$   &-0.00183 &-0.00183 &-0.00183
         &$\gamma(\lambda_5)$   & 0.0062  & 0.0312  & 0.0435 \\
6 &(2,2,4)       &$\mu_6$       &-0.00016 &-0.00015 &-0.00014
         &$\gamma(\mu_6)$       & 0.5894  & 0.6875  & 0.7383 \\
  &              &$\nu_6$       & 0.00026 & 0.00026 & 0.00026
         &$\gamma(\nu_6)$       & 0.0120  & 0.0023  &-0.0024 \\
  &              &$\delta_6$    &-0.00003 &-0.00004 &-0.00004
         &$\gamma(\delta_6)$    & & & \\
  &              &$\gamma_6$    &-0.00012 &-0.00012 &-0.00011
         &$\gamma(\gamma_6)$    & 0.1326  & 0.1893  & 0.2179 \\
  &              &$\lambda_6$   & 0.00004 & 0.00003 & 0.00002
         &$\gamma(\lambda_6)$   & & & \\
7 &(1,1,5)       &$\mu_7$       & 0.00009 & 0.00009 & 0.00010
         &$\gamma(\mu_7)$       & & & \\
  &              &$\nu_7$       & 0.00001 & 0.00001 & 0.00001
         &$\gamma(\nu_7)$       & & & \\
  &              &$\delta_7$    & 0.00000 &-0.00001 &-0.00001
         &$\gamma(\delta_7)$    & & & \\
  &              &$\lambda_7$   &-0.00001 &-0.00001 &-0.00001
         &$\gamma(\lambda_7)$   & & & \\
  &(\=3,\=3,\=3) &$\alpha_7$    & 0.00006 & 0.00006 & 0.00005
         &$\gamma(\alpha_7)$    & & & \\
  &              &$\beta_7$     & 0.00000 & 0.00000 & 0.00000
         &$\gamma(\beta_7)$     & & & \\
8 &(4,4,0)$^*$   &$\mu_8$       &-0.00002 & 0.00000 & 0.00000
         &$\gamma(\mu_8)$       & & & \\
  &              &$\nu_8$       &-0.00013 &-0.00011 &-0.00011
         &$\gamma(\nu_8)$       & 0.9582  & 0.9260  & 0.8957 \\
  &              &$\delta_8$    & 0.00027 & 0.00025 & 0.00024
         &$\gamma(\delta_8)$    & 0.7319  & 0.7755  & 0.7940 \\
  &              &$\lambda_8$   & 0.00107 & 0.00105 & 0.00104
         &$\gamma(\lambda_8)$   & 0.1332  & 0.1423  & 0.1465 \\
\hline
13&(5,5,1)$^*$   &$\mu_{13}$    &-0.00005 &-0.00004 &-0.00004
         &$\gamma(\mu_{13})$    & & & \\
  &              &$\nu_{13}$    & 0.00001 & 0.00001 & 0.00002
         &$\gamma(\nu_{13})$    & & & \\
  &              &$\delta_{13}$ & 0.00013 & 0.00011 & 0.00011
         &$\gamma(\delta_{13})$ & 0.9835  & 1.0328  & 1.0499 \\
  &              &$\lambda_{13}$&-0.00059 &-0.00058 &-0.00057
         &$\gamma(\lambda_{13})$& 0.2270  & 0.2052  & 0.1932 \\
18&(6,6,0)$^*$   &$\mu_{18}$    &-0.00001 & 0.00000 & 0.00000
         &$\gamma(\mu_{18})$    & & & \\
  &              &$\nu_{18}$    &-0.00006 &-0.00005 &-0.00005
         &$\gamma(\nu_{18})$    & & & \\
  &              &$\delta_{18}$ & 0.00006 & 0.00005 & 0.00005
         &$\gamma(\delta_{18})$ & & & \\
  &              &$\lambda_{18}$& 0.00034 & 0.00033 & 0.00033
         &$\gamma(\lambda_{18})$& 0.2448  & 0.1989  & 0.1748 \\
25&(7,7,1)$^*$   &$\mu_{25}$    &-0.00002 &-0.00001 &-0.00001
         &$\gamma(\mu_{25})$    & & & \\
  &              &$\nu_{25}$    & 0.00001 & 0.00002 & 0.00002
         &$\gamma(\nu_{25})$    & & & \\
  &              &$\delta_{25}$ & 0.00004 & 0.00004 & 0.00004
         &$\gamma(\delta_{25})$ & & & \\
  &              &$\lambda_{25}$&-0.00021 &-0.00020 &-0.00020
         &$\gamma(\lambda_{25})$& 0.3781  & 0.3210  & 0.2892 \\
\end{tabular}
\end{table}

\begin{table}
\caption{Phonon frequencies of TA($X$) mode and associated mode Gr\"uneisen
parameters for silicon at three different volumes, corresponding to
lattice constants $a$ of 10.00, 10.18, and 10.26 Bohr respectively.
The reference value (r.v.) is obtained by taking into account
interactions with all the atoms included in the real space box defined
by our sampling of the Brillouin zone (10 special points).
The other values are obtained by limiting the interactions to the successive
nearest neighbors ($NN$) shells.
The phonon frequencies are expressed in cm$^{-1}$.}
\label{tab:phonX}
\begin{tabular}{crrrrrr}
     & &$\omega_{TA(X)}$ &
     & &$\gamma_{TA(X)}$ & \\
$NN$ &$a$=10.00  &$a$=10.18  &$a$=10.26
     &$a$=10.00  &$a$=10.18  &$a$=10.26 \\
\hline
0    &383.462  &361.688  &352.426
     &  1.080  &  1.102  &  1.107 \\
1    &196.622  &206.021  &209.256
     & -1.016  & -0.727  & -0.598 \\
2    &148.056  &161.684  &166.347
     & -1.961  & -1.340  & -1.082 \\
3    &135.107  &150.376  &155.576
     & -2.413  & -1.609  & -1.287 \\
4    &131.783  &147.576  &152.949
     & -2.560  & -1.695  & -1.352 \\
5    &106.607  &128.275  &135.460
     & -4.389  & -2.633  & -2.016 \\
6    &105.327  &127.507  &134.829
     & -4.557  & -2.705  & -2.060 \\
7    &105.390  &127.438  &134.725
     & -4.525  & -2.692  & -2.053 \\
8    &122.392  &142.025  &148.656
     & -3.438  & -2.180  & -1.711 \\
13   &128.111  &146.863  &153.222
     & -3.131  & -2.019  & -1.595 \\
18   &122.443  &142.249  &148.906
     & -3.473  & -2.189  & -1.710 \\
25   &119.535  &140.618  &147.431
     & -3.690  & -2.280  & -1.762 \\
\hline
r.v. &119.903  &140.466  &147.347
     & -3.690  & -2.295  & -1.782 \\
\end{tabular}
\end{table}

\begin{table}
\caption{Phonon frequencies of TA($L$) mode and associated mode Gr\"uneisen
parameters for silicon at three different volumes, corresponding to
lattice constants $a$ of 10.00, 10.18, and 10.26 Bohr respectively.
The reference value (r.v.) is obtained by taking into account
interactions with all the atoms included in the real space box defined
by our sampling of the Brillouin zone (10 special points).
The other values are obtained by limiting the interactions to the successive
nearest neighbors ($NN$) shells.
The phonon frequencies are expressed in cm$^{-1}$.}
\label{tab:phonL}
\begin{tabular}{crrrrrr}
     & &$\omega_{TA(L)}$ &
     & &$\gamma_{TA(L)}$ & \\
$NN$ &$a$=10.00  &$a$=10.18  &$a$=10.26
     &$a$=10.00  &$a$=10.18  &$a$=10.26 \\
\hline
0    &383.462  &361.688  &352.426
     &  1.080  &  1.102  &  1.107 \\
1    &147.432  &151.460  &152.695
     & -0.609  & -0.396  & -0.295 \\
2    &119.157  &127.508  &130.363
     & -1.494  & -1.041  & -0.845 \\
3    &120.881  &129.204  &132.089
     & -1.459  & -1.033  & -0.848 \\
4    &124.496  &132.393  &135.122
     & -1.346  & -0.954  & -0.783 \\
5    & 90.610  &103.339  &107.667
     & -3.001  & -1.951  & -1.547 \\
6    & 79.247  & 93.544  & 98.304
     & -3.889  & -2.390  & -1.845 \\
7    & 76.339  & 91.167  & 96.078
     & -4.197  & -2.534  & -1.942 \\
8    & 98.491  &110.644  &114.798
     & -2.630  & -1.745  & -1.396 \\
13   &101.679  &113.235  &117.210
     & -2.416  & -1.628  & -1.312 \\
18   & 96.809  &108.985  &113.149
     & -2.680  & -1.776  & -1.420 \\
25   & 96.631  &108.918  &113.746
     & -2.532  & -1.959  & -1.737 \\
\hline
r.v. & 96.239  &108.626  &112.867
     & -2.742  & -1.814  & -1.451 \\
\end{tabular}
\end{table}

\begin{table}
\caption{Mean square atomic displacement $B$ of silicon atoms
at $T$=295~K for three different volumes
(lattice constants $a$ of 10.00, 10.18, and 10.26 Bohr),
and the corresponding experimental data at $a$=10.26~Bohr.
The lower part of the table presents a band-by-band decomposition of $B$.
The values are expressed in \AA$^{2}$.}
\label{tab:thm-fact}
\begin{tabular}{ccc|cccc|ccccc}
& & & \multicolumn{3}{c}{Present work}
  & & \multicolumn{4}{c}{Experimental data (a=10.26)} &\\
& & & a=10.00 & a=10.18 & a=10.26 &
  & Ref.~\cite{Aldred-Hart73} & Ref.~\cite{Graf81}
  & Ref.~\cite{Fehlmann79}    & Ref.~\cite{Krec-Steiner84} &\\
\hline
&$B$&  & 0.4967 & 0.4745 & 0.4707 & & 0.4613 & 0.4500 & 0.4515 & 0.4660 & \\
\hline
&$B$(TA)    &  & 0.2298 & 0.2052 & 0.1992 & & & & & & \\
&$B$(LA$_1$)&  & 0.1592 & 0.1510 & 0.1482 & & & & & & \\
&$B$(LA$_2$)&  & 0.0496 & 0.0543 & 0.0563 & & & & & & \\
&$B$(LO+TO) &  & 0.0582 & 0.0641 & 0.0672 & & & & & & \\
\end{tabular}
\end{table}

% \end{multicols}
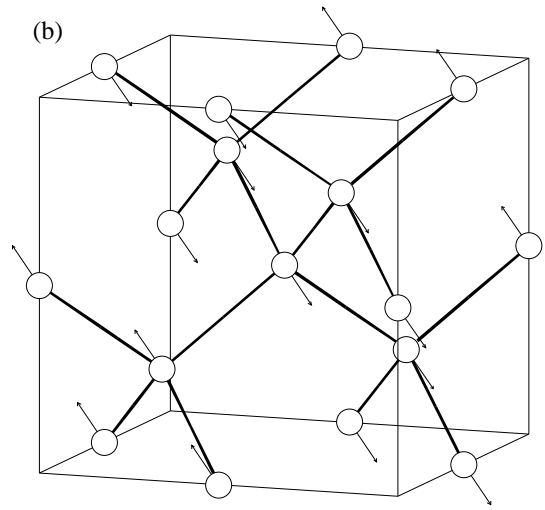
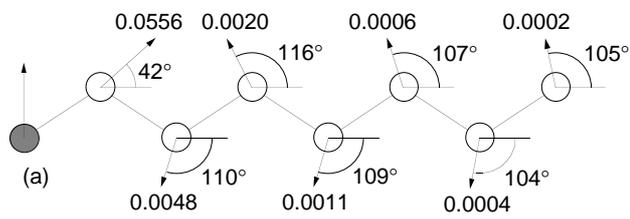
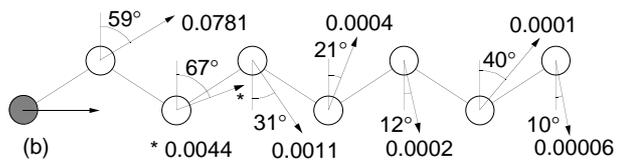
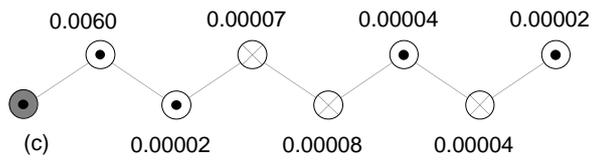
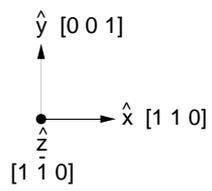
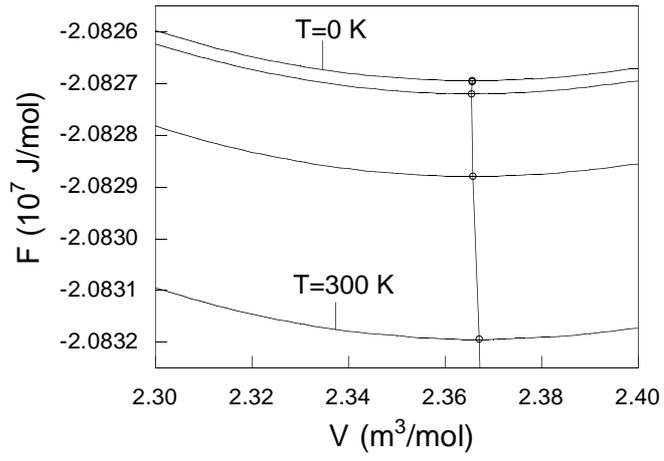
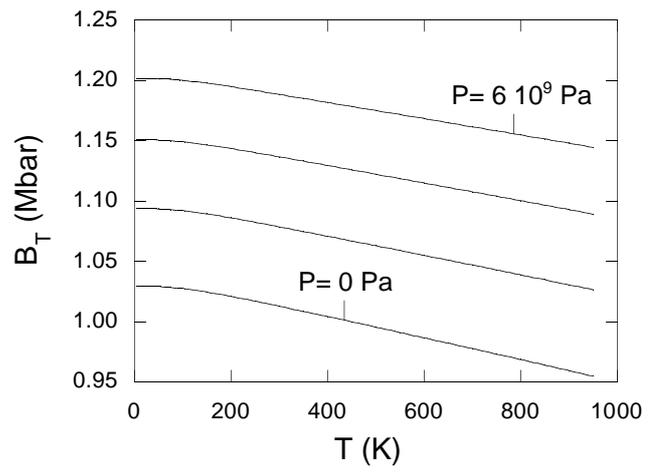
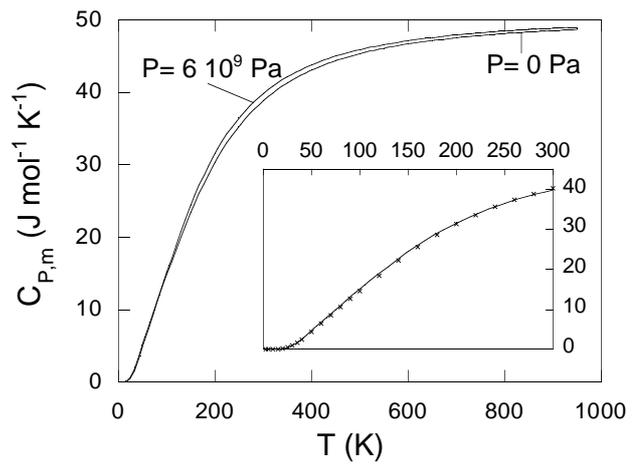
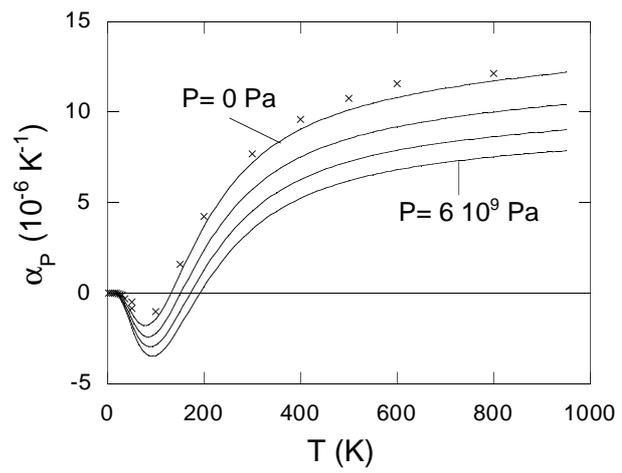
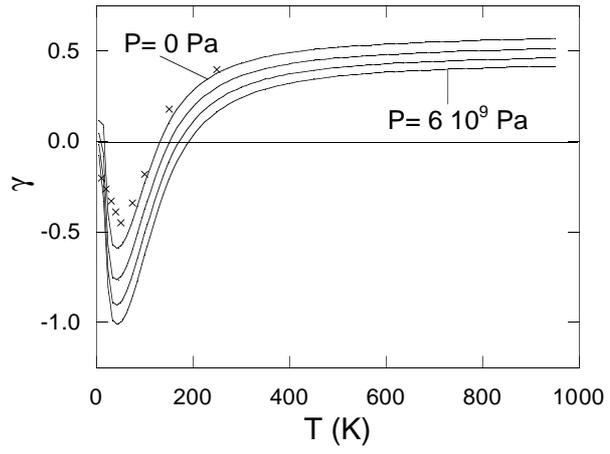
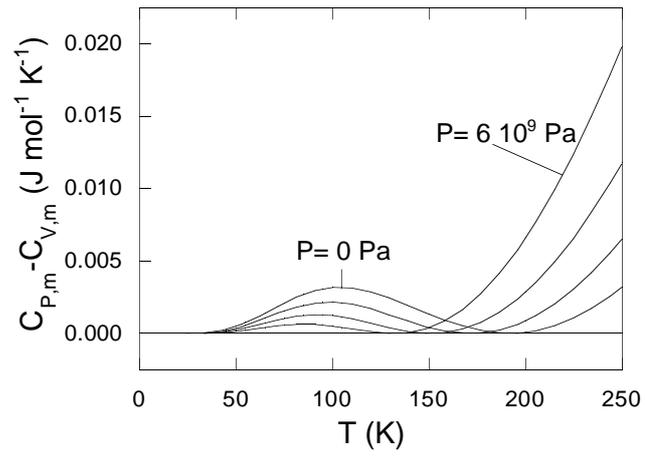
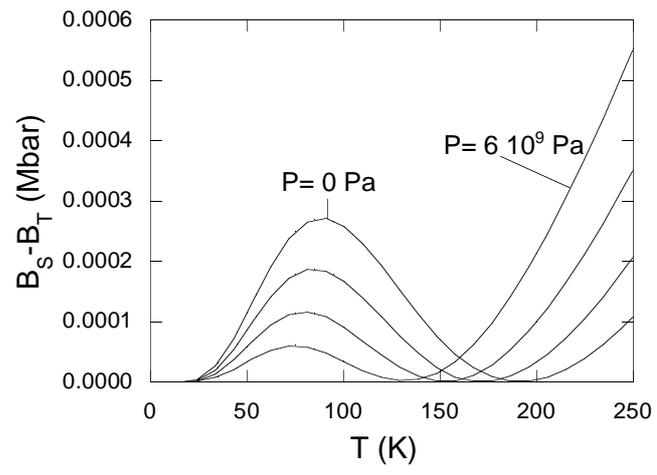
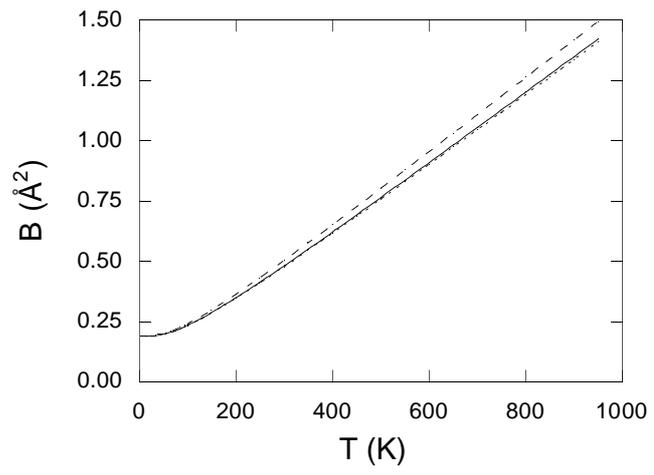
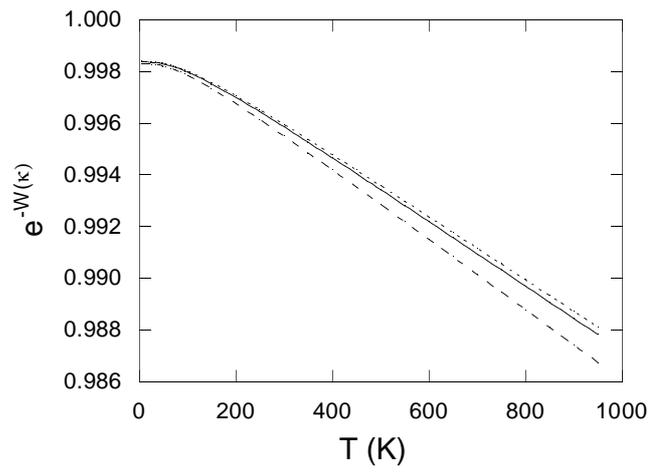
\end{document}